\documentclass{PoS}
\usepackage{lineno}

\title{Non-identical particle femtoscopy in Pb$-$Pb collisions at $
\mathbf{\sqrt{{\textit s}_{\rm NN}}}=2.76$ TeV
measured with ALICE}

\ShortTitle{Non-identical particle femtoscopy in Pb$-$Pb collisions at $
\mathbf{\sqrt{{\textit s}_{\rm NN}}}=2.76$ TeV
measured with ALICE}

\author{\speaker{Ashutosh Kumar Pandey (for the ALICE Collaboration)}
\\
       Indian Institute of Technology Bombay, Mumbai, India\\
        E-mail: \email{ashutosh.kumar.pandey@cern.ch}}

\abstract{Two-particle femtoscopic correlations between non-identical charged particles for different charge combinations are measured in Pb-Pb collisions at $\sqrt{s_{\rm NN}}$ = 2.76 TeV with ALICE at the LHC. The three-dimensional two-particle correlation functions are studied in different centrality bins. The femtoscopic source size parameter ($R_{Out}$) and emission asymmetry ($\mu$) are extracted. It is observed that the average source size of the system and emission asymmetry between particles increase from peripheral to central events.}

\FullConference{XXXIX International Conference on High Energy Physics (ICHEP 2018)\\
		4-11 July 2018\\
		Coex, Seoul}

\begin{document}
\section{Non-identical particle femtoscopy}
\label{nonidfemto} Femtoscopic techniques, which analyze the momentum correlations of
produced particles at small relative momenta, are used to study the
space$-$time characteristics of the created system. Due to Final State Interactions (FSI) among the particles, the two-particle correlations for non-identical pairs
are sensitive to space$-$time coordinates of the particle emission points
as well as the difference in average emission points (emission asymmetry) of different particle species.


\section{Method}
\label{method}
The experimental correlation function is constructed as $C({\bf k^*})
= N({\bf k^*})/D({\bf k^*})$ where $k^*$ is the momentum of the first particle in
the Pair Rest Frame (PRF), $N({\bf p}_{\rm a},{\bf p}_{\rm b})$ and
$D({\bf p}_{\rm a},{\bf p}_{\rm b})$ are the distributions when both particles coming from the same event and from two different events, respectively. 

The Koonin-Pratt equation \cite{prattkoonineqn} is given by
\begin{eqnarray}
 C({\bf{k^*}}) = \int d{\bf{r'}} |\psi({\bf{k^*}},{\bf{r'}})| ^{2} S(\bf{r'}) .
\end{eqnarray}
where $\psi({\bf{k^*}},{\bf{r'}})$ is the pair wave function which contains all the interactions
between both particles of pair.
The emission point spatial distribution was parametrized by the following functional form:

\begin{eqnarray}
 S({\bf{r}}) = exp \left( - \frac{(r_{\rm out} - \mu_{\rm out})^2}{R_{\rm out}^2} - \frac{r_{\rm side} ^2}{R_{\rm side}^2} - \frac{r_{\rm long} ^2}{R_{\rm long}^2} \right)
\end{eqnarray}
where $R_{\rm out}$, $R_{\rm side}$ and $R_{\rm long}$ are three different sizes in Out, Side and Long directions \cite{OSL}, respectively, with the mean
value $\mu_{\rm out}$ corresponding to the emission asymmetry, $r_{\rm out}$, $r_{\rm side}$ and $r_{\rm long}$ are the components of the relative separation vector {\bf{r}} of the emission points.

\section{Analysis details}
The present measurements are based on the study of pion-kaon femtoscopic correlations in Pb$-$Pb collisions measured at
$\mathbf{\sqrt{{\textit s}_{\rm NN}}}=2.76$ TeV by the ALICE detector \cite{alice} in 2011. The analysis has been performed for
central, semi-central and peripheral collisions determined by the forward V0 detector. The charged tracks were reconstructed
using the TPC detector only. Tracks with a transverse momentum within 0.19 GeV/$c$ <
$p_{\rm T}$ < 1.5 GeV/$c$ measured in the pseudo-rapidity range $|\eta| < 0.8$ are
selected. Combined information from TPC and TOF is used to
identify charged tracks as pions and kaons.
The uncorrelated pair background is constructed by pairing tracks from 
different events in same trigger class. 

\section{Results}

Using the mentioned source function as in Eq.(2.2), one can numerically integrate
Eq.(2.1) with the corresponding wave function (indicating the
type of interactions) to calculate the correlation function.
The calculated correlation is compared to
the measured via a $\chi^{2}$ test.
In this work, the CorrFit software package \cite{corrfit} was used to perform the 
numerical fitting described above and extract the experimental
${R_{\rm out}}$ and $\mu_{\rm out}$.

\begin{figure}
\centering
\includegraphics[scale=0.35]{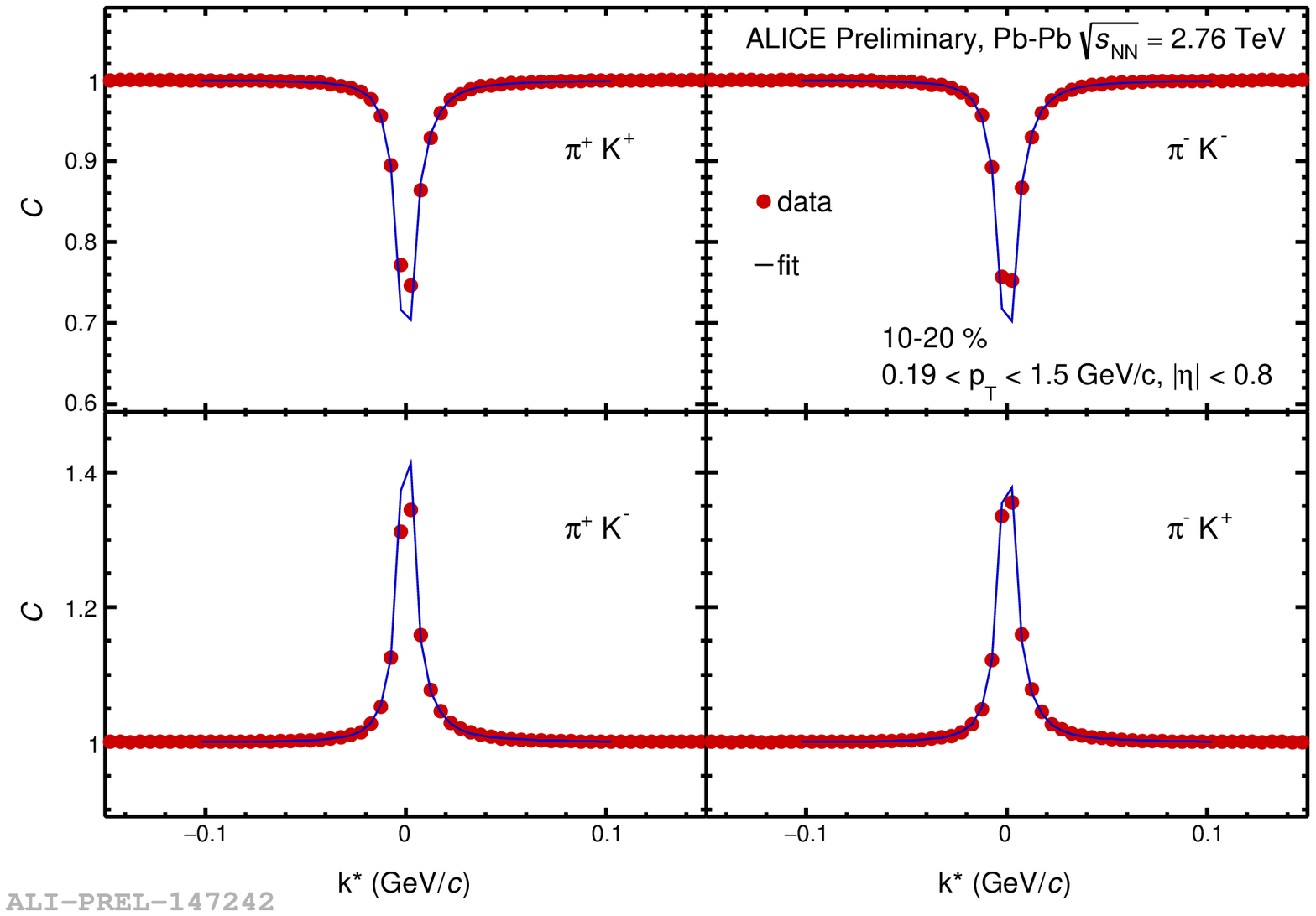}
\includegraphics[scale=0.35]{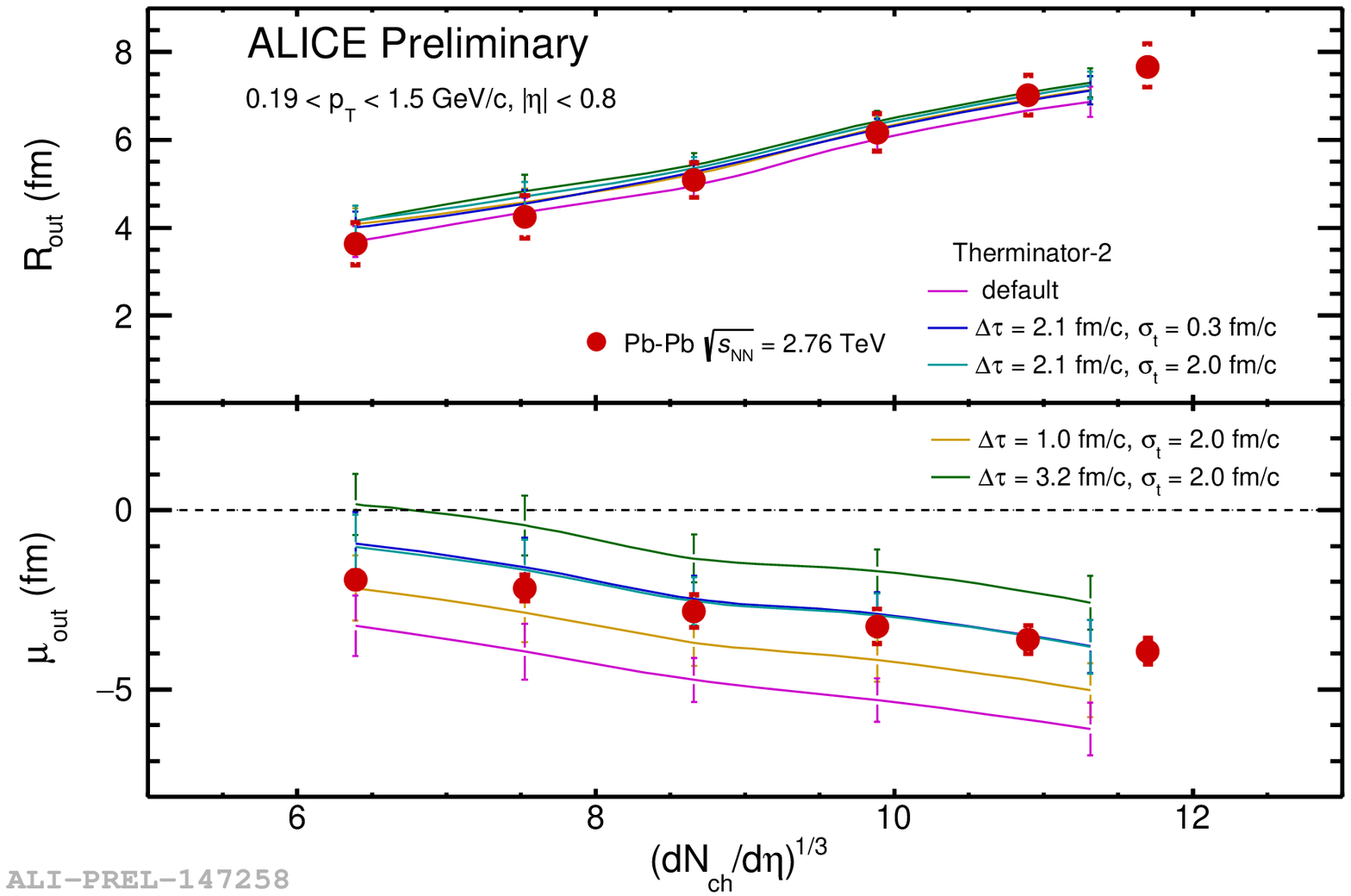}
\caption{Left: The pion-kaon correlation function for all charge combinations with their fits; Right: Source size (upper panel) and pion-
kaon emission asymmetry (lower panel) from pion-kaon correlation functions for Pb$-$Pb collisions at $\mathbf{\sqrt{{\textit s}_{\rm NN}}}=2.76$ TeV
as a function of $(dN_{ch}/d\eta)^{1/3}$ .}
\label{fig3}
\end{figure}

The right plot in Fig. \ref{fig3} shows the pion-kaon source size and emission
asymmetry as a function of the cube root of charged particle multiplicity 
density  in Pb$-$Pb collisions at $\mathbf{\sqrt{{\textit s}_{\rm NN}}}=2.76$ TeV. One observes that the system size and the extracted emission asymmetry increase with 
event multiplicity. This implies that pions
are emitted closer to the centre of the source.
The results are compared to the predictions from the
Therminator2 model \cite{therm2} and it was found that an introduction of a time delay of 2.1 fm/$c$
in kaon emission time, the result \cite{adam_noniden} is in good agreement with
the experimental measurement.
\section{Conclusion}
The first measurements of pion-kaon femtoscopic correlations in Pb$-$Pb 
collisions at $\mathbf{\sqrt{{\textit s}_{\rm NN}}}=2.76$ TeV have been performed. The radius of the source ${R_{\rm out}}$ and an observed finite emission asymmetry show a decreasing trend from central to peripheral
collisions. The results are consistent with the Therminator2 coupled with (3+1)D viscous hydrodynamic
model calculations of pion-kaon 
emission asymmetry when an additional time delay of 2.1 fm/$c$ is
introduced for the kaons.


\begin{thebibliography}{99}
\medskip
\bibitem{prattkoonineqn} S.E. Koonin, PLB70 (1977) {\bf{43}}, S. Pratt {\it et al.}, PRC42 (1990) 2646
\bibitem{OSL}Michael Annan Lisa,  Scott Pratt,  Ron Soltz,  Urs Wiedemann, {\bf  Annual Review of Nuclear and Particle Science }, Vol. 55:357-402 (2005)
\bibitem{alice} K. Aamodt {\it et al}. The ALICE experiment at the CERN LHC. JINST, 3:S08002, 2008.
\bibitem{corrfit}A. ~Kisiel, NUKLEONIKA, {\bf 49} , S81-S83, (2004)
\bibitem{therm2} M. ~Chojnacki, A.~Kisiel, W.~Florkowski and W. ~Broniowski, Comput. Phys. Commun. {\bf 183},746 (2012).  
\bibitem{adam_noniden} A. Kisiel, {\bf arXiv:1804.06781}, (2018). 
\end{thebibliography}
\end{document}